\documentclass[12pt,times]{article}

 \usepackage[pdftex]{color,graphicx}
\usepackage{enumerate}
\usepackage{amsmath}
\usepackage{amssymb}
\usepackage{alltt}
\usepackage{setspace}
\usepackage{subfigure}
\usepackage{sidecap}
\usepackage{sectsty}
\partfont{\small}
\sectionfont{\small}
\usepackage[left=2cm,top=2cm,right=3cm,head=1cm,foot=1cm]{geometry}
\numberwithin{equation}{section}
\let\oldsqrt\sqrt
\def\sqrt{\mathpalette\DHLhksqrt}
\def\DHLhksqrt#1#2{%
\setbox0=\hbox{$#1\oldsqrt{#2\,}$}\dimen0=\ht0
\advance\dimen0-0.2\ht0
\setbox2=\hbox{\vrule height\ht0 depth -\dimen0}%
{\box0\lower0.4pt\box2}}
\newcommand{\al}{\alpha}

\newcommand{\ta}{\theta}

\newcommand{\ph}{\varphi}
\newcommand{\da}{\dagger}

\newcommand{\M}{\mathcal M}

\newcommand{\ml}{\left(\begin{matrix}}
\newcommand{\mr}{\end{matrix}\right)}

\newcommand{\del}{\delta}
\newcommand{\Del}{\Delta}

\newcommand{\ka}{\kappa}
\newcommand{\VPMNS}{V_{\text{PMNS}}}

\newcommand{\im}{\text{Im}}
\newcommand{\sgn}{\text{sgn}}
\begin{document}
\begin{center}\textbf{Neutrino Mass Matrices with $M_{ee} = 0$}\end{center}
\begin{center} Yoni BenTov$^1$ and A. Zee$^{1,2}$\end{center}
\textit{$^1$\,Department of Physics, University of California, Santa Barbara CA 93106}\\
\textit{$^2$\,Kavli Institute for Theoretical Physics, University of California, Santa Barbara CA 93106}
\begin{abstract}
Motivated by the possibility that the amplitude for neutrinoless double beta decay may be much smaller than the planned sensitivity of future experiments, we study ansatze for the neutrino mass matrix with $M_{ee} = 0$. For the case in which CP is conserved, we consider two classes of real-valued mass matrices: ``Class I" defined by $|M_{e\mu}| = |M_{e\tau}|$, and ``Class II" defined by $|M_{\mu\mu}| = |M_{\tau\tau}|$. The important phenomenological distinction between the two is that Class I permits only ``small" values of $V_{e3}$ up to $\sim 0.03$, while Class II admits ``large" values of $V_{e3}$ up to its empirical upper limit of $0.22$. Then we introduce CP-violating complex phases into the mass matrix. We show that it is possible to have tribimaximal mixing with $M_{ee} = 0$ and $|M_{\mu\tau}| = |M_{\mu\mu}| = |M_{\tau\tau}|$ if the Majorana phase angles are $\pm\pi/4$. Alternatively, for smaller values of $|M_{\mu\tau}| = |M_{\mu\mu}| = |M_{\tau\tau}|$ it is possible to obtain $|V_{e3}| \sim 0.2$ and generate relatively large CP-violating amplitudes. To eliminate phase redundancy, we emphasize rephasing any mass matrix with $M_{ee} = 0$ into a standard form with two complex phases.  The discussion alternates between analytical and numerical but remains purely phenomenological, without any attempt to derive mass matrices from a fundamental theory.
\end{abstract}
\section{Data and Conventions}\label{intro}
The present empirical knowledge of neutrino oscillations can be summarized qualitatively as follows \cite{rev1,rev2}. We observe a deficiency of electron neutrinos originating from the sun and attribute this to oscillations described roughly by a mixing angle $\ta_{\text{solar}} \sim 0.5-0.8$ and a mass-squared difference $\Del m_{\text{solar}}^2 \sim 10^{-5}-10^{-4}$ eV$^2$. We also observe a deficiency of muon neutrinos in the earth's atmosphere from incident cosmic rays and attribute this to oscillations described roughly by a mixing angle $\ta_{\text{atm}} \sim 0.6-1$ and a mass-squared difference $\Del m_{\text{atm}}^2 \sim 10^{-3}$ eV$^2$. The commonly accepted theoretical interpretation of the data is that all three flavors of neutrinos -- $\nu_e,\,\nu_\mu$ and $\nu_\tau$ -- participate in oscillations. In this work we base our quantitative empirical understanding of three-flavor neutrino oscillations on the analysis of Gonzalez-Garcia and Maltoni \cite{rev2}, who report the entries of the 3-by-3 neutrino mixing matrix $V$ as having magnitudes\footnote{There are varying degrees of confidence levels assigned to the different fits in the review. To impose as little theoretical prejudice as possible, we will always quote the 3$\sigma$ bounds, which are the least restrictive.}
\begin{equation}\label{eq:Vexp}
|V_{\text{exp}}| \approx \ml 0.77\!\!-\!0.86&0.50\!\!-\!0.63&0.00\!\!-\!0.22\\0.22\!\!-\!0.56&0.44\!\!-\!0.73&0.57\!\!-\!0.80\\0.21\!\!-\!0.55&0.40\!\!-\!0.71&0.59\!\!-\!0.82 \mr\;,
\end{equation}
where the bounds are correlated such that $V$ is unitary. We also quote the recently updated report by Gonzalez-Garcia, Maltoni and Salvado \cite{rev3} for the angles\footnote{The report quotes two sets of ranges for the angles, depending on uncertainties in the capture cross section of gallium. The distinction between the two sets is a slight change in the range of $\ta_3$ and in the upper bound of $\ta_2$, with $\ta_1$ and the mass-squared splittings unaffected. We take the least restrictive bounds whenever possible. Also, the notation in the reference is $\ta_{23} \equiv \ta_1,\, \ta_{13} \equiv \ta_2$ and $\ta_{12} \equiv \ta_3$.}
\[ 0.620 \leq \ta_1 \leq 0.934\;,\;\; 0.00\leq \ta_2 \leq 0.218\;,\;\;  0.550\leq \ta_3 \leq 0.658 \]
and the mass-squared differences
\begin{align*}
&m_2^2-m_1^2 = 7.59 \ml +0.61\\ -0.69 \mr\times 10^{-5}\;\text{eV}^2\;,\;\;\text{and}\\
&m_3^2-m_1^2 = \left\{ \begin{matrix}+2.46\pm0.37\times 10^{-3}\;\text{eV}^2\;\;\text{(``normal hierarchy")}\\ -2.36\pm0.37\times 10^{-3}\;\text{eV}^2\;\;\text{(``inverted hierarchy")}\end{matrix} \right.\;\;.
\end{align*}
Although we know the mass-squared differences $m_{ij}^2 \equiv m_i^2-m_j^2$, we do not know the actual value of any of the $m_i$. Thus to compare with oscillation data, we compute the ratio of mass-squared differences
\begin{equation}\label{eq:Rexp}
R \equiv \frac{m_3^2-m_1^2}{m_2^2-m_1^2} = \left\{ \begin{matrix} +25.5\,\text{ to }+41.0\;\;\text{(``normal" hierarchy)}\\ -39.6\,\text{ to }-24.3\;\;\text{(``inverted" hierarchy)} \end{matrix} \right.\;\;.
\end{equation}
To study the mixing matrix, we use the standard angular parameterization \cite{angles} for unitary matrices given by 
\begin{equation}\label{eq:KPM}
V = \mathcal K\VPMNS\M \;,\; \text{ where }\; \mathcal K \equiv \text{diag}(e^{\,i\ka_1},e^{\,i\ka_2},e^{\,i\ka_3})\; ,\;\; \M \equiv \text{diag}(e^{\,i\rho},e^{\,i\sigma},1)
\end{equation}
and
\begin{align}\label{eq:PMNS}
\VPMNS &\equiv \ml 1&0&0\\0&c_1&s_1\\ 0&s_1&-c_1 \mr \ml c_2&0&\hat s_2^*\\0&1&0\\-\hat s_2&0&c_2 \mr \ml -c_3&s_3&0\\s_3&c_3&0\\0&0&1 \mr \nonumber \\
&= \ml -c_2c_3&c_2s_3&\hat s_2^*\\ c_1s_3+s_1\hat s_2c_3&c_1c_3-s_1\hat s_2s_3&s_1c_2\\ s_1s_3-c_1\hat s_2c_3 & s_1c_3+c_1\hat s_2s_3 & -c_1c_2 \mr\;.
\end{align}
Here $c_I \equiv \cos\ta_I$, $s_I \equiv \sin\ta_I$ and $\hat s_2 \equiv s_2\,e^{\,i\delta_{\text{CP}}}$, and we have chosen the sign conventions in $\VPMNS$ to minimize the number of minus signs that appear. The angles in $\mathcal K$ are unphysical and can be chosen arbitrarily.
\\\\
We will assume that the neutrinos are Majorana. In this case the Majorana phase matrix $\M$ is physically meaningful and contributes to the amplitude for neutrinoless double beta decay. In addition, the neutrino mass matrix $M$ is symmetric. We will work in the basis for which the charged lepton mass matrix is diagonal with real positive entries, called the flavor basis. In this basis the neutrino mass matrix is 
\begin{equation}\label{eq:M}
M = V^*DV^\da
\end{equation}
where $D \equiv \text{diag}(m_1,m_2,m_3)$. Here $m_i \geq 0$ denote the physical masses of the three neutrinos.
\\\\
There is some degree of rephasing freedom in the neutrino mass matrix $M$, and we will return to this point in a later section on CP violation. For now we simply wish to clarify a potential source of confusion for the case in which CP is conserved. If CP is conserved, then $M$ can be taken as real, and we can without loss of generality set $\del_{\text{CP}} = 0$. However, we cannot set $\rho$ and $\sigma$ equal to zero, since the Majorana phase matrix $\M$ appears squared in the mass matrix $M$. With $\mathcal K = I$ and $\del_{\text{CP}} = 0$, we have $M = \mathcal K^*\VPMNS^*\M^*D\M^*\VPMNS^\da\mathcal K^* = \VPMNS\tilde D\,\VPMNS^T$, where we have defined the diagonal matrix
\begin{equation}\label{eq:Dtilde}
\tilde D \equiv \ml \tilde m_1\\&\tilde m_2\\&&m_3 \mr\;\text{ with}\;\; \tilde m_1 \equiv m_1\,e^{-i2\rho}\;\text{and}\;\;\tilde m_2 \equiv m_2\,e^{-i2\sigma}\;.\end{equation}
\\
The notation is such that $m_{1,2}$ are real and positive while $\tilde m_{1,2}$ are complex. (Also, with our phase conventions $m_3$ is always real and positive.) Thus the choice of $0$ or $\frac{\pi}{2}$ for $\rho$ and $\sigma$ generates non-removable minus signs associated with $m_1$ and $m_2$, which yield qualitatively different textures for the mass matrix $M$. In the CP-conserving case, it is convenient to separate these signs from $V$ and instead associate them with the diagonal matrix $\tilde D$.
\section{Neutrinoless Double Beta Decay and $M_{ee} = 0$}\label{Mee}
As discussed at the end of the previous section, the choice of signs in $\tilde m_{1,2}$ imply qualitatively different textures for the mass matrix. To motivate a particular choice, we recall the well-known fact that a direct way to measure one of the entries in $M$ is in neutrinoless double beta decay, the amplitude of which is proportional to $|M_{ee}|$. From Eqs. (\ref{eq:KPM}-\ref{eq:M}) we have
\begin{equation}\label{eq:Mee}
|M_{ee}| = \left| c_2^2\left( c_3^2\,e^{\,i\al_1}m_1+s_3^2\,e^{\,i\al_2}m_2 \right)+s_2^2m_3\right|
\end{equation}
where $\al_1 \equiv -2(\rho+\del_{\text{CP}})$ and $\al_2 \equiv -2(\sigma+\del_{\text{CP}})$. Thus in general, $|M_{ee}|$ depends on all three masses $m_1,m_2$ and $m_3$, the two angles $\ta_2$ and $\ta_3$, and two phases $\al_1$ and $\al_2$.
\\\\
A brief review of the current status of neutrinoless double beta decay was given recently by Bilenky \cite{beta}, which we now summarize\footnote{An early review of neutrinoless double beta decay was given by Zel'dovich and Klhopov \cite{betadecay}.}. The Heidelberg-Moscow and CUORICINO experiments imply the upper bounds $|M_{ee}| \leq (0.3-1.2)$ eV and $|M_{ee}| \leq (0.3-1.7)$ eV, respectively. The future experiments CUORE, EXO, GENIUS and MAJORANA plan to significantly improve the sensitivity to roughly $|M_{ee}| \sim (1-7)\times10^{-2}$ eV. 
\\\\
These values should be understood in comparison to the $m_{ij}^2$ data above Eq. (\ref{eq:Rexp}), which imply 
\[8.3\times10^{-3}\text{eV} \leq \sqrt{m_{21}^2} \leq 9.1\times10^{-3}\text{eV} \text{ and } 4.6\times10^{-2}\text{eV}\leq \sqrt{m_{31}^2}\leq 5.3\times10^{-2}\text{eV}\]
for the normal hierarchy $m_1 < m_2 < m_3$. 
\\\\
If $m_3 \gg m_{1,2}$, then $m_3 \approx \sqrt{m_{31}^2} \sim 5\times10^{-2}$ eV, but since $s_2^2 \leq 4.68\times 10^{-2}$ the large $m_3$ is suppressed by the small $s_2^2$ in $|M_{ee}|$.  Thus $|M_{ee}|$ is at most $\sim 10^{-3}$ eV, which is an order of magnitude smaller than the planned sensitivity of future experiments. If $m_3 > m_{1,2}$ but all three masses are still almost equal, then the $m_3$ term drops out and $|M_{ee}| \approx m_1c_2^2|c_3^2+s_3^2\,e^{\,i(\al_2-\al_1)}|$. With the bounds given below Eq. (\ref{eq:Vexp}), this implies $0.26 \leq |M_{ee}|/m_1 \leq 1$, where the upper bound occurs for $\al_2=\al_1$ and $\ta_2 = 0$.
\\\\
Thus for any normal hierarchy, $|M_{ee}|$ tends to be smaller than the other entries in $M$. Using this as guidance, we suppose that $M_{ee}$ could be tiny and thereby set $M_{ee} = 0$. In other words, throughout this paper we assume that the amplitude for neutrinoless double beta decay is zero, at least as a leading order approximation \cite{meezero, meezero1, meezero2, meezero3}.
\section{Tribimaximal Mixing with $M_{ee} = 0$}\label{tribi}
As has been noted independently by many authors \cite{wtribi, tribi}, the theoretical ansatz of ``tribimaximal mixing" defined as
\begin{equation}\label{eq:TB}
\VPMNS = V_{\text{TB}} \equiv \ml \frac{-2}{\sqrt6}&\frac{1}{\sqrt3}&0\\ \frac{1}{\sqrt6}&\frac{1}{\sqrt3}&\frac{1}{\sqrt2}\\ \frac{1}{\sqrt6}&\frac{1}{\sqrt3}&\frac{-1}{\sqrt2} \mr \approx \ml -0.82&0.58&0\\ 0.41&0.58&0.71\\ 0.41&0.58&-0.71 \mr
\end{equation}
is compatible with the empirical bounds given in $|V_{\text{exp}}|$. If neutrino oscillations conserve CP, then we can write the neutrino mass matrix in the flavor basis as $M = \VPMNS \tilde D\VPMNS^T$ , where $\tilde D \equiv \text{diag}(\tilde m_1,\tilde m_2,m_3)$ and $\tilde m_i \equiv \pm m_i$ with uncorrelated signs. We can thereby define a ``tribimaximal mass matrix" $M_{\text{TB}} \equiv V_{\text{TB}}\,\tilde D\,V_{\text{TB}}^T$ associated with the ansatz of tribimaximal mixing. Explicitly, this mass matrix reads\footnote{Since the mass matrix is symmetric, we display explicitly only its upper triangle.}
\begin{equation}\label{eq:MTB}
M_{\text{TB}} = \frac{1}{3}\left[ \tilde m_1\ml 2&-1&-1\\ &1/2&1/2\\&&1/2 \mr+\tilde m_2\ml 1&1&1\\&1&1\\&&1 \mr \right]+\frac{m_3}{2}\ml 0&0&0\\&1&-1\\&&1 \mr\;.
\end{equation}
For all values of $\tilde m_i$ and $m_3$, this matrix exhibits the symmetry $M_{e\mu} = M_{e\tau}$ and $M_{\mu\mu} = M_{\tau\tau}$ \cite{tribisym, 23sym}. We stress that although the condition $V = V_{\text{TB}}$ necessarily implies $M_{e\mu} = M_{e\tau}$ and $M_{\mu\mu} = M_{\tau\tau}$, the converse is not true: $M_{e\mu} = M_{e\tau}$ and $M_{\mu\mu} = M_{\tau\tau}$ do not necessarily imply tribimaximal mixing. 
\\\\
Two appealing examples of tribimaximal mass matrices with $M_{ee} = 0$ are obtained by choosing the values $(\tilde m_1,\tilde m_2,m_3) = (-1,2,9)$ and $(-1,2,11)$, which give
\begin{equation}\label{eq:MTB1}
M_{\text{TB}}^{(-1,+2,\,9)} = m_\nu \ml 0&1&1\\&5&-4\\&&5 \mr  \;\;\;\text{ and }\;\;\; M_{\text{TB}}^{(-1,+2,11)} = m_\nu\ml 0&1&1\\&6&-5\\&&6 \mr
\end{equation}
respectively. The first has $R \approx 27$, while the second has $R = 40$ exactly, which correspond nearly to the lower and upper empirical bounds for $R$. 
\\\\
In the mass matrix $M_{\text{TB}}$ the sign flip $(\tilde m_1,\tilde m_2) \to (-\tilde m_1,-\tilde m_2)$ effects the exchange 
\begin{equation}
(M_{\mu\mu},M_{\mu\tau}) \to -(M_{\mu\tau},M_{\mu\mu})\;.
\end{equation}
This means that given one tribimaximal mass matrix, we can always find a second tribimaximal mass matrix by interchanging the magnitudes of $M_{\mu\mu}$ and $M_{\mu\tau}$.
\\\\
Thus from (\ref{eq:MTB1}) we can immediately write the matrices\footnote{We have used the rephasing freedom in $M$ to move around the minus signs. See Section~\ref{phases}.}
\begin{equation}\label{eq:MTB2}
M_{\text{TB}}^{(+1,-2,9)} = m_\nu \ml 0&1&1\\&4&-5\\&&4 \mr\;\;\;\text{ and }\;\;\; M_{\text{TB}}^{(+1,-2,11)} = m_\nu\ml 0&1&1\\&5&-6\\&&5 \mr
\end{equation}
which also predict $R \approx 27$ and $R = 40$ respectively. 
\\\\
A mass matrix that resembles the examples given above but with non-tribimaximal mixing is\footnote{Many authors have proposed parametrizations of deviations from tribimaximal mixing \cite{nontribi}.}
\begin{equation}\label{eq:MnTB}
M_{\text{nTB}} \equiv m_\nu\ml \frac{1}{5}&1&1\\&5&-5\\&&3  \mr \implies R \approx 29\;\;\text{ and }\;\;|\VPMNS| \approx \left(
\begin{array}{ccc}
 0.84 & 0.54 & 0.02 \\
 0.33 & 0.54 & 0.77 \\
 0.43 & 0.65 & 0.63
\end{array}
\right)
\end{equation}
\\
which was suggested in the context of a particular model \cite{porto}. Since $M_{\mu\mu} \neq M_{\tau\tau}$ the resulting mixing matrix is not tribimaximal, as can be seen from the nonzero $V_{e3}$. On the other hand, both $M_{\text{nTB}}$ and $M_{\text{TB}}$ share the property $M_{e\mu} = M_{e\tau}$. The matrix $M_{\text{nTB}}$ also has $M_{ee}$ smaller than the other entries.
\\\\
In an attempt to systematically study this distinction, we consider the ``Class I" ansatz 
\begin{equation}\label{eq:ClassI}
M_{\text{I}} \equiv m_\nu\ml 0&1&1\\&M_{\mu\mu}&M_{\mu\tau}\\ &&M_{\tau\tau} \mr \qquad\qquad\rm{(Class\;I)}
\end{equation}
with $M_{ee} = 0$ and $M_{e\mu} = M_{e\tau}$. To further classify deviations from tribimaximal mixing with $M_{ee} = 0$, we also consider the ``Class II" ansatz\footnote{In Class II, the value $M_{\mu\mu} = M_{\tau\tau} = 5$ is merely a convenient normalization for comparing the empirically allowed mass matrices with those of Class I.}
\begin{equation}\label{eq:ClassII}
M_{\text{II}} \equiv m_\nu \ml 0&M_{e\mu}&M_{e\tau}\\&5&M_{\mu\tau}\\&&5 \mr \qquad\qquad\rm{(Class\;II)}
\end{equation}
with $M_{\mu\mu} = M_{\tau\tau}$. Since oscillation experiments cannot determine the overall scale of $M$, we from now on set $m_\nu = 1$ and treat the entries of $M$ as dimensionless numbers. 
\\\\
We emphasize to the reader that we make no attempt to derive these mass matrices from any theoretical model but instead study these matrices on purely phenomenological grounds.
\section{Rephasing the Mass Matrix}\label{phases}
Before proceeding to study the matrices $M_{\text{I}}$ and $M_{\text{II}}$, we should comment on the significance of various signs that may appear in the mass matrix. Consider the most general 3-by-3 complex symmetric matrix $M$ with $M_{ee} = 0$:
\begin{equation}\label{eq:Mee0}
M = \ml 0&a_{e\mu}\,e^{\,i\ph_{e\mu}}&a_{e\tau}\,e^{\,i\ph_{e\tau}}\\&a_{\mu\mu}\,e^{\,i\ph_{\mu\mu}}&a_{\mu\tau}\,e^{\,i\ph_{\mu\tau}}\\&&a_{\tau\tau}\,e^{\,i\ph_{\tau\tau}} \mr
\end{equation}
where $a_{\al\beta}$ and $\ph_{\al\beta}$ are real numbers. Using the form $M = V^*D V^\da$ with $V = \mathcal K\VPMNS \M$ introduced in Section~\ref{intro}, we have $M = \mathcal K^* \hat M \mathcal K^*$, where $\hat M = \VPMNS^*\M^*D\M^*\VPMNS^\da$, and thus $\hat M = \mathcal K M\mathcal K$. We are free to choose the phases in $\mathcal K$ as we please, since they are unphysical. Choosing $\ka_1 = \frac{1}{2}\ph_{\mu\mu}-\ph_{e\mu},\,\ka_2 = -\frac{1}{2}\ph_{\mu\mu}$ and $\ka_3 = -\frac{1}{2}\ph_{\tau\tau}$ gives
\begin{equation}\label{eq:Mee0rephased}
\hat M = \ml 0&a_{e\mu}&a_{e\tau}\,e^{\,i\ph}\\ &a_{\mu\mu}&a_{\mu\tau}\,e^{\,i\eta}\\ &&a_{\tau\tau} \mr
\end{equation}
where $\ph \equiv \ph_{e\tau}-\ph_{e\mu}+\frac{1}{2}\ph_{\mu\mu}-\frac{1}{2}\ph_{\tau\tau}$ and $\eta \equiv \ph_{\mu\tau}-\frac{1}{2}\ph_{\mu\mu}-\frac{1}{2}\ph_{\tau\tau}$. We may thus dispense with the matrix $M$ and consider only the matrix $\hat M$. Henceforth when there is no risk of confusion we put any mass matrix $M$ into the form of $\hat M$ and then drop the hat for notational convenience.
\\\\
For the case in which $M$ is real, the phases reduce to the signs $\pm 1$. The above argument shows that any real-valued neutrino mass matrix with $M_{ee} = 0$ can be put into the form
\begin{equation}\label{eq:Mee0rephasedreal}
M = \ml 0&|M_{e\mu}|&\zeta |M_{e\tau}|\\ &|M_{\mu\mu}|&\zeta' |M_{\mu\tau}|\\ &&|M_{\tau\tau}| \mr
\end{equation}
where each of $\zeta$ and $\zeta'$ can be either $+1$ or $-1$. The matrix $M$ can be multiplied on both sides by the matrix $Z \equiv \text{diag}(1,1,-1)$, which transforms $(\zeta,\zeta') \to (-\zeta,-\zeta')$ and thereby leaves the product $\zeta\zeta'$ unchanged. Since $\det(ZMZ) = \det M$, all observables based on the $M$ in (\ref{eq:Mee0rephasedreal}) are invariant under $M \to ZMZ$ and therefore depend only on $\sgn(M_{e\tau}M_{\mu\tau})$, not on $\zeta = \sgn(M_{e\tau})$ and $\zeta' = \sgn(M_{\mu\tau})$ individually. 
\\\\
If we allow $M_{\mu\tau}$ to range over all real numbers, then in both Classes I and II we can take all other entries in $M$ to be strictly non-negative. Given this choice, it will turn out furthermore that only $M_{\mu\tau} < 0$ can fit data. This can be seen from the form of $M_{\text{tribi}}$ with $m_3 \gg m_{1,2}$.
\\\\
To summarize, we will first study real-valued mass matrices of Classes I and II given in Eqs. (\ref{eq:ClassI}) and (\ref{eq:ClassII})
with $M_{ee} = 0$, $M_{\mu\tau} < 0$ and all other entries positive. 
\section{Analytic Preliminaries}\label{analytic}
\begin{figure}[h]
\begin{center}
\fbox{
	\begin{minipage}{14 cm}
	\begin{center}
		\includegraphics[scale=0.65]{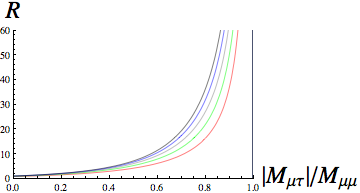}
	\end{center}
	\vspace{-10pt}
	\caption{\small{The mass-squared difference ratio $R \equiv m_{31}^2/m_{21}^2$ for the $\mu\tau$-symmetric ansatz Eq. (\ref{eq:mutau}). The curves correspond to the fixed values $M_{\mu\mu}$ = 3 (Red), 4 (Green), 5 (Gray), 6 (Blue), 7 (Black). For $|M_{\mu\tau}|/M_{\mu\mu} > 1$, reflect the graph about the vertical line $|M_{\mu\tau}|/M_{\mu\mu} = 1$}}
	\label{Classmtsym-R}	
	\end{minipage}
	}
\end{center}
\end{figure}

\begin{figure}[h]
\begin{center}
\fbox{
	\begin{minipage}{16 cm}
		\subfigure[$~|V_{e1}|\text{ vs. }|M_{\mu\tau}|/M_{\mu\mu}$]
		{
		\includegraphics[scale=0.60]{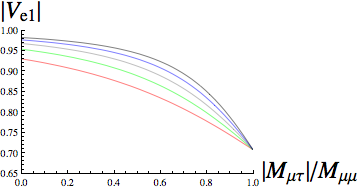}
		\label{Classmtsym-Ve1}
		}
		\subfigure[$~V_{e2}\text{ vs. }|M_{\mu\tau}|/M_{\mu\mu}$]
		{
		\includegraphics[scale=0.60]{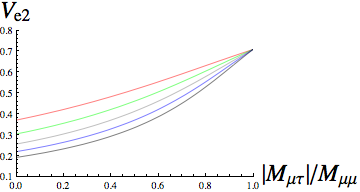}
		\label{Classmtsym-Ve2}
		}
		\vspace{-10pt}
		\caption{\small{The elements $V_{e1}$ and $V_{e2}$ of the mixing matrix for the $\mu\tau$-symmetric ansatz Eq. (\ref{eq:mutau}). The curves correspond to the fixed values $M_{\mu\mu}$ = 3 (Red), 4 (Green), 5 (Gray), 6 (Blue), 7 (Black). For $|M_{\mu\tau}|/M_{\mu\mu} > 1$, reflect each graph about the vertical line $|M_{\mu\tau}|/M_{\mu\mu} = 1$.} }
	\end{minipage}
	}
\end{center}
\end{figure}

If CP is conserved in the neutrino sector, there are 6 potential observables in neutrino phenomenology: 3 angles $\ta_i$ and 3 masses $m_i$. Accordingly, a general 3-by-3 real symmetric matrix has 6 independent parameters and thereby makes no predictions. By fixing $M_{ee} = 0$ we impose a constraint and thus fix one of the parameters \cite{porto}, namely the angle $\ta_2$ (and thus $V_{e3}$), according to the relation 
\begin{equation}\label{eq:Mee0constraint}
\tan^2\ta_2 = -\left(\frac{c_3^2\tilde m_1+ s_3^2\tilde m_2}{m_3}\right)\;.
\end{equation}
Empirically we know that $\ta_2 \leq 0.22$, so $\tan^2\ta_2 \ll 1$. This tells us that we cannot have $m_3 \ll m_1 \sim m_2$, thus forbidding the inverted hierarchy\footnote{In more detail, the inverted case $m_3 < m_1 < m_2$ would require roughly $|\frac{m_2}{m_1}\tan^2\ta_3\pm1| < 6\times10^{-2}$, where the $\pm$ is fixed according to $\tilde m_1\tilde m_2 > 0$ (plus sign) or $\tilde m_1\tilde m_2 < 0$ (minus sign). For the $+$ case, this inequality is clearly impossible to satisfy since 1 is larger than $10^{-2}$. For the minus case, the ratio $m_2/m_1$ would have to be of order 1 but fine-tuned to two decimal places. We will not consider this particular case and thereby specialize to $m_1 < m_2 < m_3$.} for the ansatz $M_{ee} = 0$ \cite{glashow}. 
\\\\
As a limiting case, for $\ta_2 \to 0$ we predict $m_1 \to m_2\,\tan^2\ta_3$ and thus fix all three neutrino masses. Since $0.36 \leq \tan^2\ta_3 \leq 0.60$ and $6.9\times10^{-5}\,\text{eV}^2 \leq m_2^2-m_1^2 \leq 8.2\times10^{-5}\,\text{eV}^2$, we have
\[\lim_{\ta_2\to0}m_1 = \sqrt{\frac{m_2^2-m_1^2}{\cot^4\ta_3-1}} = 6.4\times10^{-3}\,\text{eV}\;\text{ to }\;1.1\times10^{-2}\,\text{eV}\]
and
\[\lim_{\ta_2\to0}m_3 = 4.6\times10^{-2}\,\text{eV}\;\text{ to }\;5.4\times10^{-2}\,\text{eV}\;.\]
\\
Note that these ranges are rather narrow: $m_1$ can be only as large as $11/6.4 \sim 1.7$ of its minimum value, and $m_3$ can be only as large as $54/46 \sim 1.2$ of its minimum value. 
\\\\
After fixing $M_{ee} = 0$, the next step is to specialize either to Class I  by imposing $M_{e\mu} = M_{e\tau}$ or to Class II by imposing $M_{\mu\mu} = M_{\tau\tau}$. Either choice will fix the angle $\ta_1$ in terms of the other parameters, thus reducing the number of free parameters to four: the three masses $m_i$ and the angle $\ta_3$.
\\\\
If we were to impose the condition $M_{e\mu} = M_{e\tau}$ (Class I), then this would fix
\[\tan\ta_1 = \frac{1-xs_2}{1+xs_2}\;\;\;\text{ where }\;\;\; x \equiv \frac{c_3^2\tilde m_1+s_3^2\tilde m_2-m_3}{c_3s_3(\tilde m_2-\tilde m_1)}\;.\]
Instead, if we were to impose the condition $M_{\mu\mu} = M_{\tau\tau}$ (Class II), then this would fix\footnote{This condition results in a quadratic equation for $\tan\ta_1$ whose two roots are $\tan \ta_1 = -y \pm \sqrt{y^2+1}$. We choose the $+$ sign to keep $\tan\ta_1$ positive.} $\tan \ta_1 = -y+ \sqrt{y^2+1}$ where
\[ y \equiv \frac{(\tilde m_1-\tilde m_2)s_2\sin(2\ta_3)}{c_2^2m_3-p\,\tilde m_1-q\,\tilde m_2}\;\;\;\text{ with }\;\;\;p \equiv s_3^2-s_2^2c_3^2\;\;\;\text{ and }\;\;\;q \equiv c_3^2-s_2^2s_3^2\;.\]
In the limit $\ta_2 \to 0$, the two conditions become equivalent and imply $\ta_1 \to \pi/4$. Therefore the limiting case $\ta_2 \to 0$ of our matrices of Classes I and II corresponds to the $\mu\tau$-symmetric ansatz
\begin{equation}\label{eq:mutau}
M_{\mu\tau-\text{sym}} \equiv \ml 0&1&1\\&a&b\\&&a \mr
\end{equation}
\\
which along with the possibility of $\ta_1 = \pi/4$ was studied by many authors \cite{23sym}. 
\\\\
At this point we should comment on $\mu\tau$ symmetry in the neutrino mass matrix. Since the mass of the $\tau$ is an order of magnitude larger than the mass of the muon, the effective Lagrangian at energy scales below $m_\tau$ already exhibits deviations from any underlying $\mu\tau$ symmetry that may exist at high energy. At the energy scale of neutrino masses $m_\nu \ll m_\mu$, any high-energy $\mu\tau$ symmetry should be badly broken and thus corrections to $M$ are to be expected in general. Thus the $\mu\tau$-symmetric texture of (\ref{eq:mutau}) should be thought of at most as a useful starting point for a phenomenological analysis.
\\\\
It is also worth remarking that tribimaximal mixing implies $M_{e\mu} = M_{e\tau}$ and $M_{\mu\mu} = M_{\tau\tau}$, but the converse is not true. For the $\mu\tau$-symmetric ansatz (\ref{eq:mutau}), the third eigenvector is $V_{\al3} \propto (0,1,-1)$, exactly as for tribimaximal mixing, irrespective of the values of $a$ and $b$. However, the other two eigenvectors are only proportional to $(-2,1,1)$ and $(1,1,1)$ for the particular case $b = 1-a$. For example, $a = 5$ and $b = 1-5 = -4$ reproduce the matrix $M_{\text{TB}}^{-1,+2,9}$ of Eq. (\ref{eq:MTB1}). On the other hand, changing $b$ to $-4.2$ gives
\[M = \ml 0&1&1\\&5&-4.2\\&&5 \mr \implies R \approx 36\;\;\text{ and }\;\;|\VPMNS| \approx \left(
\begin{array}{ccc}
 0.80 & 0.60 & 0 \\
 0.43 & 0.56 & 0.71 \\
 0.43 & 0.56 & 0.71
\end{array}
\right)\]
which deviates from tribimaximal mixing in the first two columns of $V$. Attempting to increase $|M_{\mu\tau}|$ to $4.3$ would result in $R \approx 42$, just above the upper limit, but the resulting $V$ would remain compatible with the bounds in Eq. (\ref{eq:Vexp}).
\\\\
Attempting instead to decrease $|M_{\mu\tau}|$ below 4 results in a mass-squared difference ratio that is too small. For example, $|M_{\mu\tau}| = 3.9$ would result in $R \approx 23$, which is less than the empirical lower bound of 25.5. As for the previous case, the resulting $V$ would be compatible with data. 
\\\\
Thus in the $\mu\tau$-symmetric case, the experimental constraint on $R = m_{31}^2/m_{21}^2$ is more stringent than the constraints for the entries in $V$. This can be understood from Fig.~\ref{Classmtsym-R}, in which we plot $R$ as a function of $|M_{\mu\tau}|/M_{\mu\mu}$ for the matrix in Eq. (\ref{eq:mutau}). As $|M_{\mu\tau}|/M_{\mu\mu} \to 1$ we have $m_{21}^2 \to 0$, so that the ratio $R$ diverges as $|M_{\mu\tau}| \to M_{\mu\mu}$, which we will discuss in more detail in the next section (see Eq. (\ref{eq:MB})).
\pagebreak
\\
In comparison, there is no divergence in either $V_{e1}$ or $V_{e2}$, whose sensitivity to the ratio $|M_{\mu\tau}|/M_{\mu\mu}$ is displayed in Figs.~\ref{Classmtsym-Ve1} and~\ref{Classmtsym-Ve2}.
\\
Having gained an analytic understanding of the mass matrices in Class I (\ref{eq:ClassI}) and Class II (\ref{eq:ClassII}), as well as their $\mu\tau$-symmetric intersection (\ref{eq:mutau}), we now turn to numerics. The analysis that follows should be useful for classifying perturbations away from tribimaximal mixing within the $\mu\tau$-symmetric ansatz as well as for classifying deviations from $\mu\tau$ symmetry in more general mass matrices.
\section{Real Mass Matrices: Class I ($M_{e\mu} = M_{e\tau}$)}\label{real1}
\begin{figure}[h]
\begin{center}
\fbox{
	\begin{minipage}{16 cm}
		\centering
		\subfigure[$~M_{\tau\tau}\text{ vs. }M_{\mu\mu}\text{ for all allowed values of }|M_{\mu\tau}|$]
		{
		\includegraphics[scale=0.50]{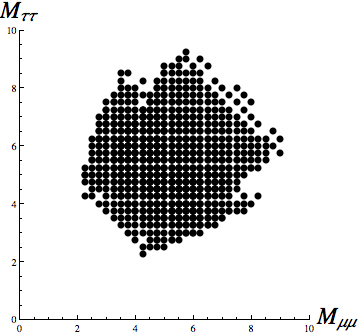}
		\label{ClassI-MmmOverMtt-MttVsMmm}
		}\hspace{20pt}
		\subfigure[$~|M_{\mu\tau}|\text{ vs. }M_{\mu\mu}\text{ with }M_{\tau\tau} = M_{\mu\mu}$]
		{
		\includegraphics[scale=0.50]{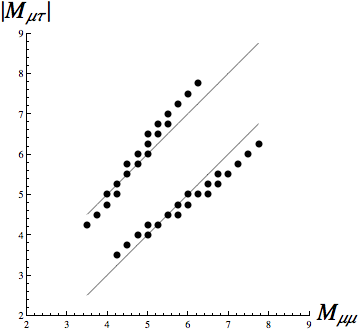}
		\label{ClassI-MmmOverMtt-MmtVsMmm}
		}
		\vspace{-10pt}
		\caption{\small{The ratio $M_{\mu\mu}/M_{\tau\tau}$ in Class I ($M_{e\mu} = M_{e\tau} = 1$). Figure 1(a) shows the values of the diagonal entries in Class I for all possible allowed values of $M_{\mu\tau}$. Figure 1(b) shows the allowed values for $M_{\mu\tau}$ for matrices of Class I that also satisfy $M_{\mu\mu} = M_{\tau\tau}$ and therefore are $\mu\tau$-symmetric. As discussed in Section~\ref{analytic}, mass matrices with $M_{\mu\mu} = M_{\tau\tau}$ for which $|M_{\mu\tau}| = M_{\mu\mu}-1$ (lower solid line) and $|M_{\mu\tau}| = M_{\mu\tau}+1$ (upper solid line) yield tribimaximal mixing.} }
	\end{minipage}
	}
\end{center}
\end{figure}

\begin{figure}[h]
\begin{center}
\fbox{
	\begin{minipage}{15 cm}
	\vspace{-5pt}
		\subfigure[$~|M_{\mu\tau}|\text{ vs. }M_{\mu\mu}\text{ for all allowed }M_{\tau\tau}$]
		{
		\includegraphics[scale=0.50]{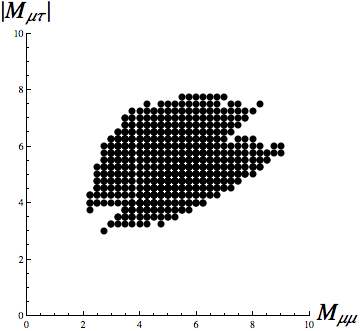}
		\label{ClassI-MmtMmm}
		}
		\subfigure[$~|M_{\mu\tau}|\text{ vs. }M_{\tau\tau}\text{ for all allowed }M_{\mu\mu}$]
		{
		\includegraphics[scale=0.50]{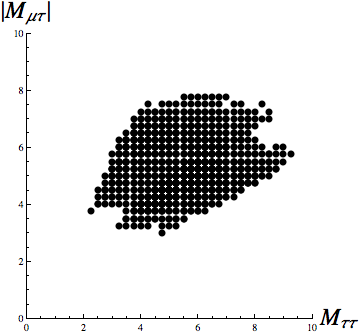}
		\label{ClassI-MmtMtt}
		}
		\vspace{-10pt}
		\caption{\small{The allowed values of $|M_{\mu\tau}|$ in Class I ($M_{e\mu} = M_{e\tau} = 1$).} }
	\end{minipage}
	}
\end{center}
\end{figure}

\begin{figure}[h]
\begin{center}
\fbox{
	\begin{minipage}{15 cm}
		\subfigure[$~V_{e3}\text{ vs. }M_{\mu\mu}\text{ for all allowed }M_{\tau\tau}\text{ and }|M_{\mu\tau}|$]
		{
		\includegraphics[scale=0.55]{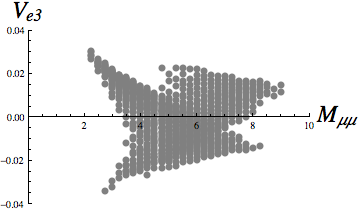}
		\label{ClassI-Ve3Mmm}
		}
		\subfigure[$~V_{e3}\text{ vs. }M_{\tau\tau}\text{ for all allowed }M_{\mu\mu}\text{ and }|M_{\mu\tau}|$]
		{
		\includegraphics[scale=0.55]{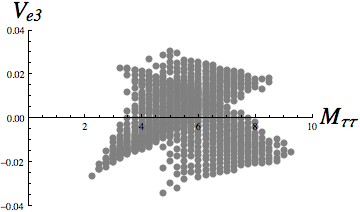}
		\label{ClassI-Ve3Mtt}
		}
		\caption{\small{$V_{e3}$ as a function of $M_{\al\beta}$ in Class I ($M_{e\mu} = M_{e\tau} = 1$). In both plots, the variables not displayed explicitly on the axes are allowed to range over all of their possible values that result in an acceptable mixing matrix $V$ and mass-squared difference ratio $R$.}}
	\end{minipage}
	}
\end{center}
\end{figure}

We now begin a numerical study of the Class I ansatz defined by Eq. (\ref{eq:ClassI}), which for the convenience of the reader we display again:
\[M_{\text{I}} \equiv \ml 0&1&1\\&M_{\mu\mu}&M_{\mu\tau}\\ &&M_{\tau\tau} \mr\]
Here $M_{\mu\tau}$ is strictly negative, and all other nonzero entries are strictly positive. 
\\\\
Figure~\ref{ClassI-MmmOverMtt-MttVsMmm} shows a plot of the allowed values for the ratio $M_{\mu\mu}/M_{\tau\tau}$ while letting $M_{\mu\tau}$ range over all its possible values. We find that the nonzero diagonal entries can lie in the ranges\footnote{Here and throughout the rest of the paper, we use the ``$\sim$" symbol to denote a rough guide for the values of the entries in $M$, to be compared with either $M_{e\mu} = M_{e\tau} = 1$ (Class I) or $M_{\mu\mu} = M_{\tau\tau} = 5$ (Class II). The idea is to get a feel for what the entries in $M$ can be, and then afterwards to hunt for precise numerical values that fit data.} $M_{\mu\mu} \sim 2-9$ and $M_{\tau\tau} \sim 2-10$. The fact that these ranges are essentially the same is something we already knew, since as discussed in Section~\ref{tribi} the case $M_{\mu\mu} = M_{\tau\tau}$ is the $\mu\tau$-symmetric subcase of Class I. 
\pagebreak\\
However, looking at Figure~\ref{ClassI-MmmOverMtt-MttVsMmm} in isolation may give the misleading impression that the case $|M_{\mu\tau}| = M_{\mu\mu} = M_{\tau\tau}$ is allowed when in fact it is experimentally ruled out, as can be seen numerically in Figure~\ref{ClassI-MmmOverMtt-MmtVsMmm}. This can also be seen in Figs.~\ref{Classmtsym-R}, ~\ref{Classmtsym-Ve1} and ~\ref{Classmtsym-Ve2} when compared with the bounds given in Eqs. (\ref{eq:Vexp}) and (\ref{eq:Rexp}).
\\\\
This can also be understood analytically as follows. The mass matrix
\begin{equation}\label{eq:MB}
M_{\text{B}} \equiv \ml 0&1&1\\&a&-a\\&&a \mr
\end{equation}
implies a ``bimaximal" mixing matrix
\begin{equation}\label{eq:bimax}
\VPMNS = V_{\text{B}} \equiv \ml \frac{-1}{\sqrt2}&\frac{1}{\sqrt2}&0\\ \frac{1}{2}&\frac{1}{2}&\frac{1}{\sqrt2}\\ \frac{1}{2}&\frac{1}{2}&\frac{-1}{\sqrt2} \mr
\end{equation}
and two equal neutrino masses, both of which are incompatible with the empirically allowed ranges quoted in (\ref{eq:Vexp}) and (\ref{eq:Rexp}). 
\\\\
Figure ~\ref{ClassI-MmmOverMtt-MmtVsMmm} shows that the set of allowed mass matrices splits into two branches, with larger and smaller $|M_{\mu\tau}|$, which yield a larger and smaller $R$ respectively. For example, for fixed $M_{\mu\mu} = M_{\tau\tau} = 5$ we recover\footnote{As discussed in Section~\ref{tribi}, we also have Eq. (\ref{eq:MTB2}).} either the matrix with $M_{\mu\tau} = -4$ and $R \approx 27$ in Eq. (\ref{eq:MTB1}), or the matrix with $M_{\mu\tau} = -6$ and $R = 40$ in Eq. (\ref{eq:MTB2}).
\\\\
We can use Figures~\ref{ClassI-MmmOverMtt-MttVsMmm} and~\ref{ClassI-MmmOverMtt-MmtVsMmm} to look for examples of mass matrices with non-tribimaximal mixing. Towards the upper limit of $|M_{\mu\tau}| \sim 8$, we find
\[M = \ml 0&1&1\\&6&-7.6\\&&6 \mr \implies R \approx 35\;\;\text{ and }\;\; |\VPMNS| \approx \left(
\begin{array}{ccc}
 0.86 & 0.50 & 0 \\
 0.36 & 0.61 & 0.71 \\
 0.36 & 0.61 & 0.71
\end{array}
\right)\;.\]
Like tribimaximal mixing, this case has $M_{e\mu} = M_{e\tau}$ and $M_{\mu\mu} = M_{\tau\tau}$ with $V_{\al3} = (0,\frac{1}{\sqrt2},\frac{-1}{\sqrt2})$. Unlike tribimaximal mixing, the second column of $V$ is not proportional to $(1,1,1)$, and the first column changes accordingly to maintain orthogonality. This is all consistent with the analytic understanding of the $\mu\tau$-symmetric ansatz from Section~\ref{analytic}.
\\\\
Thus, as emphasized throughout, this is an example with $M_{\mu\mu} = M_{\tau\tau}$ but without tribimaximal mixing. In passing, we mention that increasing $|M_{\mu\tau}|$ to $7.7$ would make $|V_{e1}|$ too large and $V_{e2}$ too small with respect to the bounds given in Eq. (\ref{eq:Vexp}). The sensitivity of $V_{e1}$ and $V_{e2}$ to changes in $|M_{\mu\tau}|$ can be seen in Figs.~\ref{Classmtsym-Ve1} and~\ref{Classmtsym-Ve2}.
\\\\
Toward the lower limit of $|M_{\mu\tau}| \sim 3$, we find
\[M = \ml 0&1&1\\&3&-2.9\\&&3.9 \mr\implies R \approx 27\;\;\text{ and }\;\; |\VPMNS| \approx \left(
\begin{array}{ccc}
 0.77 & 0.64 & 0.02 \\
 0.48 & 0.59 & 0.65 \\
 0.42 & 0.49 & 0.76
\end{array}
\right)\]
which is of a similar form to the ``nTB" matrix of Eq. (\ref{eq:MnTB}), except with $M_{\tau\tau}$ larger than $M_{\mu\mu} = |M_{\mu\tau}|$. Again in passing, we point out that increasing $|M_{\mu\tau}|$ to 3 would make $|V_{e1}|$ too small and $V_{e2}$ too large. On the other hand, keeping $|M_{\mu\tau}| = 2.9$ but increasing $M_{\tau\tau}$ to 4 would result in $R \approx 25.2$, just below the experimental lower bound, while maintaining a consistent mixing matrix $V$. An example of the sensitivity of $R$ to the ratios of various entries in $M$ can be seen in Fig.~\ref{Classmtsym-R}, although the matrix above is not $\mu\tau$ symmetric.
\\\\
Numerically we find that for all allowed values for $M_{\mu\mu}$ and $M_{\tau\tau}$, the entry $|M_{\mu\tau}|$ can be in the range $\sim 3-8$. This can be seen in Figs.~\ref{ClassI-MmtMmm} and~\ref{ClassI-MmtMtt}.
\vspace{-10pt}
\\
We conclude the study of Class I with a comment on $V_{e3}$. Figures ~\ref{ClassI-Ve3Mmm} and ~\ref{ClassI-Ve3Mtt} show that these mass matrices exhibit a maximum value $|V_{e3}| \sim 0.03$, which is rather small. However, these figures also identify that having $M_{\mu\mu}$ and $M_{\tau\tau}$ both less than 3 or greater than 8 ensures a nonzero $V_{e3}$. Recall that previously we observed $M_{\mu\mu} \sim 2-9$ and $M_{\tau\tau} \sim 2-10$, so that the narrow ranges $M_{\mu\mu} \sim 2-3$ or $8-9$, and $M_{\tau\tau} \sim 2-3$ or $8-10$ are those which necessarily produce $V_{e3} \neq 0$. These ranges are correlated, so that $M_{\mu\mu} \sim 2$ with $M_{\tau\tau} \sim 10$ is not allowed. As an example, we have:
\\
\[M = \left(
\begin{array}{ccc}
 0 & 1 & 1 \\
  & 3 & -6.4 \\
  &  & 7.5
\end{array}
\right) \implies R \approx 29\;\;\text{ and }\;\; \VPMNS \approx \left(
\begin{array}{ccc}
 -0.86 & 0.51 & 0.02 \\
 0.43 & 0.70 & 0.58 \\
 0.28 & 0.50 & -0.82
\end{array}
\right)\]
and thereby generate $|V_{e3}| \sim 0.02$, as for $M_{\text{nTB}}$. 
\\\\
In summary, matrices of Class I necessarily have ``small" values of $V_{e3}$, reaching a maximum of only $\sim 0.03$. 
\\
\section{Real Mass Matrices: Class II ($M_{\mu\mu} = M_{\tau\tau}$)}\label{real2}
\begin{figure}[h]
\begin{center}
\fbox{
	\begin{minipage}{16 cm}
	\vspace{-5pt}
	\subfigure[$~|M_{\mu\tau}|\text{ vs. }M_{e\mu}\text{ for all allowed values of }M_{e\tau}$]
	{
		\includegraphics[scale=0.53]{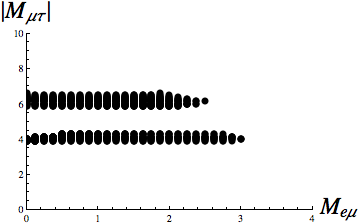}
		\label{ClassII-MmtMem}
	}\hspace{10pt}
		\subfigure[$~|M_{\mu\tau}|\text{ vs. }M_{e\tau}\text{ for all allowed values of }M_{e\mu}$]
	{
		\includegraphics[scale=0.53]{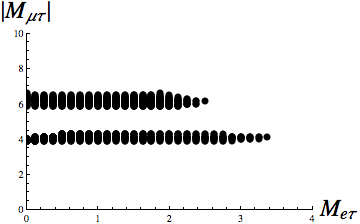}
		\label{ClassII-MmtMet}
	}
	\vspace{-10pt}
	\caption{\small{The allowed values of $|M_{\mu\tau}|$ as a function of $M_{e\mu}$ and $M_{e\tau}$ for Class II $(M_{\mu\mu} = M_{\tau\tau} = 5)$. Parameters not displayed explicitly on the axes are allowed to attain all values compatible with (\ref{eq:Vexp}) and (\ref{eq:Rexp}).}}
	\end{minipage}
	}
\end{center}
\end{figure}

\begin{figure}[h]
\centering
\fbox{
	\begin{minipage}{14 cm}
	\centering
		\includegraphics[scale=0.50]{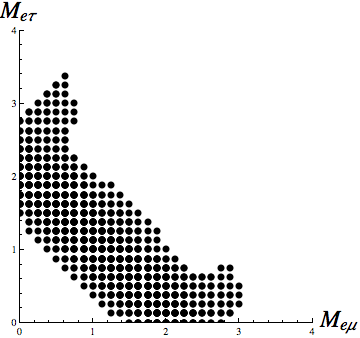}
		\vspace{-10pt}
		\caption{\small{The ratio $M_{e\tau}/M_{e\mu}$ (for all allowed $|M_{\mu\tau}|$) in Class II $(M_{\mu\mu} = M_{\tau\tau} = 5)$.}}
	\label{ClassII-MetMem}	
	\end{minipage}
	}
\end{figure}

Now consider matrices from Class II, which is defined by
\[M_{\text{II}} \equiv \ml 0&M_{e\mu}&M_{e\tau}\\&5&M_{\mu\tau}\\&&5 \mr\;.\]
In Figures ~\ref{ClassII-MmtMem} and ~\ref{ClassII-MmtMet} we plot the allowed values of $|M_{\mu\tau}|$ as a function of $M_{e\mu}$ and $M_{e\tau}$. We find that $|M_{\mu\tau}|$ is constrained to be very close to either 4 or 6, reminiscent of the tribimaximal cases given in Eqs. (\ref{eq:MTB1}) and (\ref{eq:MTB2}). This tells us that matrices in Class II necessarily exhibit near tribimaximal mixing if the ratio $M_{e\mu}/M_{e\tau}$ is close to 1. 
\\\\
We emphasize that the reason the ratio $M_{e\mu}/M_{e\tau}$ characterizes proximity to tribimaximal mixing in Class II is simply because the data in (\ref{eq:Vexp}) and (\ref{eq:Rexp}) constrain $|M_{\mu\tau}|$ to be close to 4 or 6 (in units for which $M_{\mu\mu} = M_{\tau\tau} = 5$). Otherwise, as discussed below Eq. (\ref{eq:mutau}), $\mu\tau$ symmetry does not imply tribimaximal mixing.
\\\\
In Figure~\ref{ClassII-MetMem} we examine the ratio $M_{e\mu}/M_{e\tau}$. The case $M_{e\mu}/M_{e\tau} = 1$ is allowed for $M_{e\mu} = M_{e\tau} \sim 0.8 - 1.5$, but for values out of this range for either $M_{e\mu}$ or $M_{e\tau}$, the mixing matrix will deviate significantly from the tribimaximal ansatz while still fitting data.
\\\\
In particular, it is possible for matrices of Class II ($M_{\mu\mu} = M_{\tau\tau},\,M_{e\mu} \neq M_{e\tau}$) to fit data with either $M_{e\mu} = 0$ or $M_{e\tau} = 0$ but not both. For example:
\[M = \left(
\begin{array}{ccc}
 0 & 0 & 1.4 \\
  & 5 & -4 \\
  &  & 5
\end{array}
\right)\implies R \approx 41\;\;\text{ and }\;\; |\VPMNS|\approx \left(
\begin{array}{ccc}
 0.86 & 0.50 & 0.11 \\
 0.29 & 0.66 & 0.69 \\
 0.42 & 0.56 & 0.71
\end{array}
\right)\;.\]
\\
Decreasing $M_{e\tau}$ to 1.3 would make $R$ and $|V_{e1}|$ too large and $V_{e2}$ too small with respect to the bounds in Eq. (\ref{eq:Vexp}). This two-zero texture is labeled ``Case $A_1$" in a study by Frampton, Glashow and Marfatia about possible zeros in the neutrino mass matrix in the flavor basis \cite{zeros}. The salient feature of both their work and ours is the possibility of a ``large" $V_{e3}$, on which we now elaborate.
\begin{figure}[h]
\begin{center}
\fbox{
	\begin{minipage}{15 cm}
	\subfigure[
	$~V_{e3}\text{ vs. }M_{e\mu}\text{ for all allowed }M_{e\tau}\text{ and }|M_{\mu\tau}|$]
		{
		\includegraphics[scale=0.55]{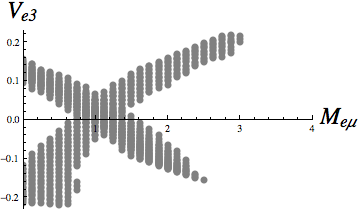}
		\label{ClassII-Ve3Mem}
		}
	\subfigure[
	$~V_{e3}\text{ vs. }M_{e\tau}\text{ for all allowed }M_{e\mu}\text{ and }|M_{\mu\tau}|$]
		{
		\includegraphics[scale=0.55]{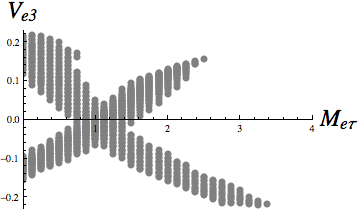}
		\label{ClassII-Ve3Met}
		}
	\vspace{-10pt}
	\caption{\small{$V_{e3}$ as a function of $M_{e\al}$ in Class II $(M_{\mu\mu} = M_{\tau\tau} = 5)$.}}
	\end{minipage}
	}
\end{center}
\end{figure}
\\\\
For either $M_{e\mu}$ or $M_{e\tau}$ less than $\sim 0.6$ or greater than $\sim 1.6$ in Class II, a nonzero $V_{e3}$ is generated. Figures ~\ref{ClassII-Ve3Mem} and ~\ref{ClassII-Ve3Met} show that the empirical upper limit $|V_{e3}| \sim 0.2$ can be generated for $M_{e\mu}$ or $M_{e\tau}$ close to $\sim 0$ or $3$. Using these along with Figure~\ref{ClassII-MetMem}, we find:
\[M = \left(
\begin{array}{ccc}
 0 & 0.2 & 3 \\
  & 5 & -4 \\
&  & 5
\end{array}
\right)\implies R \approx 33\;\;\text{ and }\;\;|\VPMNS| \approx \left(
\begin{array}{ccc}
 0.79 & 0.57 & 0.22 \\
 0.32 & 0.69 & 0.65 \\
 0.52 & 0.44 & 0.73
\end{array}
\right)\;.\]
Decreasing $M_{e\mu}$ to zero would generate $|V_{e3}| \approx 0.23$, which is too large if we believe the upper bound given in Eq. (\ref{eq:Vexp}).
\\\\
In summary, mass matrices of Class II can result in values for $|V_{e3}|$ anywhere from 0 to the empirical upper limit of $\sim 0.22$. In particular, when the ratio $M_{e\mu}/M_{e\tau}$ is greater than $\sim 2$ or less than $\sim 1/2$, a ``large" $|V_{e3}|$ is necessarily generated.
\section{CP Violation}\label{cp}
We will now allow for the possibility that the neutrino mass matrix violates CP. As discussed in Section~\ref{phases}, any 3-by-3 complex symmetric mass matrix with $M_{ee} = 0$ can be rephased into the form of Eq. (\ref{eq:Mee0rephased}), which we repeat for convenience:
\[M = \ml 0&|M_{e\mu}|&|M_{e\tau}|\,e^{\,i\ph}\\ &|M_{\mu\mu}|&|M_{\mu\tau}|\,e^{\,i\eta}\\ &&|M_{\tau\tau}| \mr\;.\] 
\\
We need to diagonalize $M$ to determine how $\ph$ and $\eta$ contribute to the CP-violating angle $\del_{\text{CP}}$ and to the Majorana phase angles $\rho$ and $\sigma$. (Recall the notation of Eq. (\ref{eq:KPM}).) The mapping of two phases $\ph$ and $\eta$ to three observables $\del_{\text{CP}},\,\rho$ and $\sigma$ is explained by the fact that $\rho$ and $\sigma$ are not independent parameters when $M_{ee} = 0$. With a complex mass matrix, the condition $M_{ee} = 0$ implies 
\begin{equation}\label{eq:Mee0constraintcomplex}
\tan^2\ta_2 = -\left(\frac{c_3^2\,e^{-i2\rho}m_1+s_3^2e^{-i2\sigma}m_2}{m_3}\right)
\end{equation}
which is the generalization of Eq. (\ref{eq:Mee0constraint}) with the possibility of $\rho$ and $\sigma$ being different from 0 or $\pi/2$. The imaginary part of this fixes $\rho$ in terms of $\sigma$ through the relation 
\begin{equation}\label{eq:Mee0constraintIm}
\frac{\sin(2\rho)}{\sin(2\sigma)} = -\tan^2\ta_3\,\frac{m_2}{m_1}
\end{equation}
so that only one of these phases is an independent parameter.
\\\\
The main result of the generalization to complex mass matrices is that nontrivial phases open up new regions for the allowed values of $|M_{\al\beta}|$.
\\\\
To understand this claim it is sufficient to specialize to the following example: Recall the matrix $M_{\text{B}}$ from Eq. (\ref{eq:MB}), which predicted $m_1 = m_2$ and bimaximal mixing with $|V_{e1}| = V_{e2}$, which are incompatible with the bounds in (\ref{eq:Vexp}) and (\ref{eq:Rexp}). Generalizing this matrix to the complex case
\begin{equation}\label{eq:MBc}
M_{\text{Bc}} \equiv \ml 0&1&e^{\,i\ph}\\&a&a\,e^{\,i\eta}\\&&a  \mr
\end{equation}
can split the degeneracy $m_1 = m_2$ and modify $|V_{e1}| = V_{e2}$ significantly enough to become compatible with oscillation data. In the next section, we will show that the matrix $M_{\text{Bc}}$ can result in tribimaximal mixing with $m_i = (1,2,9)$ and $m_i = (1,2,11)$, just as in the CP-conserving case discussed in Section~\ref{tribi}.
\section{Tribimaximal Mixing and Nonzero Majorana Phases}\label{complexTB}
\begin{figure}[h]
\begin{center}
\fbox{
	\begin{minipage}{16 cm}
	\subfigure[$~a\text{ vs. }R\text{ for tribimaximal mixing}$]
	{
		\includegraphics[scale=0.60]{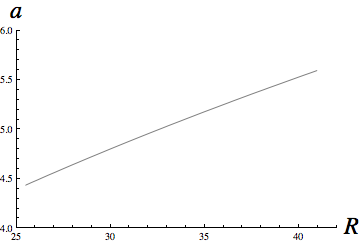}
		\label{ClassMmtEqMmm-avsR}
	}
	\subfigure[$~\eta-\pi\text{ vs. }R\text{ for tribimaximal mixing}$]
	{
		\includegraphics[scale=0.60]{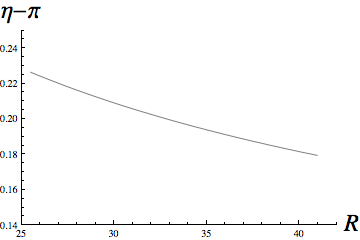}
		\label{ClassMmtEqMmm-PhipvsR}
	}
	\vspace{-10pt}
	\caption{\small{The parameters $a \equiv M_{\mu\mu} = M_{\tau\tau} = |M_{\mu\tau}|$ and $\eta \equiv \arg (M_{\mu\tau})$ in the matrix $M_{\text{Bc}}$ given in Eq. (\ref{eq:MBc}), for the particular case in which the mixing matrix is exactly tribimaximal. In this case we have $m_2 = 2m_1$ and $M_{e\mu} = M_{e\tau} = m_1$, so that setting $M_{e\mu} = M_{e\tau} = 1$ implies $m_1 = 1$, $m_2 = 2$ and thus $8.80 \leq m_3 = \sqrt{3R+1} \leq 11.1$.}}
	\end{minipage}
	}
\end{center}
\end{figure}

Consider tribimaximal mixing\footnote{Some of our work in this section overlaps with that of Z. Z. Xing \cite{meezero2}.}, meaning $\VPMNS = V_{\text{TB}}$, but with arbitrary Majorana phases so that\footnote{We remind the reader that $D = \text{diag}(m_1,m_2,m_3)$ is real and positive, and $V = \mathcal K\VPMNS \M$ is the full 3-by-3 unitary matrix including the extra phases in $\mathcal K$ and $\M$. If $\vec v_i$ denotes the $i^{\text{th}}$ column of $V$, then $M = V^*DV^\da = \sum_{i\,=\,1}m_i\vec v_i^{\,*}\vec v_i^{\,\da}$\;.} $M = V^*DV^\da$ is complex even though $\del_{\text{CP}}$ drops out since $V_{e3} = 0$. 
\\\\
For tribimaximal mixing with arbitrary Majorana phases, the condition $M_{ee} = 0$ now fixes
\begin{equation}
2m_1\,e^{-i2\rho}+m_2\,e^{-i2\sigma} = 0
\end{equation}
which corresponds to taking $\ta_2 = 0$ and $\ta_3 = \sin^{-1}(1/\sqrt3)$ in Eq. (\ref{eq:Mee0constraintcomplex}). This implies
\begin{equation}\label{eq:TBrhosigma}
m_2 = 2m_1\;\;\;\text{ and }\;\;\; \rho = \sigma+(2n-1)\frac{\pi}{2}
\end{equation}
where $n$ is any integer. Consider the case $\rho = -\sigma = -\pi/4$. Upon choosing the $\ka_i$ as given above Eq. (\ref{eq:Mee0rephased}), we find a rephased mass matrix $\hat M$ of the form
\begin{equation}\label{eq:MTBc}
\hat M = M_{\text{TBc}} \equiv \ml 0&1&1\\&a&a\,e^{\,i\eta}\\&&a \mr
\end{equation}
where
\begin{equation}\label{eq:MTBphip}
a = \frac{1}{2}\sqrt{m_3^2+1}\;\;\;\text{ and }\;\;\;\eta = \pi+\tan^{-1}\left( \frac{2m_3}{m_3^2-1} \right)\;.
\end{equation}
As displayed above, this matrix has\footnote{As discussed below Eq. (\ref{eq:Mee0rephasedreal}), we can equivalently set $\ph = \pi$ if we subtract $\pi$ from $\eta$.} $\ph = 0$. We have set the overall scale $m_\nu = m_1 = \hat M_{e\mu} = \hat M_{e\tau}$ to 1. 
\\\\
For $(m_1,m_2) = (1,2)$ we can invert the definition $R \equiv m_{31}^2/m_{21}^2$ to get $m_3 = \sqrt{3R+1}$ and thus obtain the parameters $a$ and $\eta$ as a function purely of the experimentally constrained ratio $R$:
\begin{equation}\label{eq:MTBcparameters}
a = \frac{1}{2}\sqrt{3R+2}\;\;\;\;\text{ and }\;\;\;\; \eta = \tan^{-1}\left( \frac{2}{3R}\sqrt{3R+1} \right)
\end{equation}
where $25.5 \leq R \leq 41.0$ as given in Eq. (\ref{eq:Rexp}). These parameters are plotted as a function of $R$ in Figs.~\ref{ClassMmtEqMmm-avsR} and~\ref{ClassMmtEqMmm-PhipvsR}. In units of $m_1 = 1$, the mass of the heaviest neutrino is $8.80 \leq m_3 \leq 11.1$. 
\\
Consider the two examples $m_3 = 9$ and $m_3 = 11$. For $m_i = (1,2,9)$ we have $a \approx 4.5$ and $\eta-\pi \approx 0.22$, and for $m_i = (1,2,11)$ we have $a \approx 5.5$ and $\eta-\pi \approx 0.18$. This interpolates between the two branches $|M_{\mu\tau}| \sim 4$ and $|M_{\mu\tau}| \sim 6$ of the real-valued tribimaximal mass matrix. For instance, $m_i = (1,2,10)$ (so that $R = 33$ exactly) implies $a = \frac{1}{2}\sqrt{101} \approx 5.0$ and $\eta-\pi \approx 0.20$.
\\\\
Therefore in addition to the CP-conserving case with $\ph = 0$, $\eta = \pi$ and $|M_{\mu\tau}| \neq M_{\mu\mu}$, we find a new class of allowed tribimaximal mass matrices with $\ph = 0$, $\eta \sim \pi+0.2$ and $|M_{\mu\tau}| = M_{\mu\mu}$. It is important to note that, in contrast, the case $\eta \sim 0.2$ is not allowed unless the angle $\ph$ is changed to $\sim \pi$. (Recall the notation of Eq. (\ref{eq:MBc}).) This should be understood in the context of the discussion below Eq. (\ref{eq:Mee0rephasedreal}), in which we showed that phases of $e^{\pm i\pi} = -1$ can be exchanged between $M_{e\tau}$ and $M_{\mu\tau}$.
\section{Complex Mass Matrices with $M_{e\mu} = |M_{e\tau}|$ and $M_{\mu\mu} = M_{\tau\tau} = |M_{\mu\tau}|$}
\begin{figure}[h]
\begin{center}
\fbox{
	\begin{minipage}{15 cm}
	\subfigure[$~\ph\text{ vs. }M_{\mu\mu}\text{ for all allowed values of }\eta$]
	{
		\includegraphics[scale=0.50]{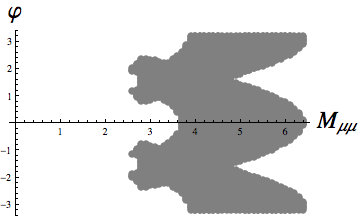}
		\label{ClassMmtEqMmm-PhiVsMmm}
	}
	\subfigure[$~\eta\text{ vs. }M_{\mu\mu}\text{ for all allowed values of }\ph$]
	{
		\includegraphics[scale=0.50]{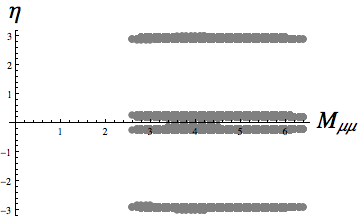}
		\label{ClassMmtEqMmm-PhipVsMmm}
	}
	\vspace{-10pt}
	\caption{\small{The phase angles as a function of $M_{\mu\mu} = M_{\tau\tau} = |M_{\mu\tau}|$ for the matrix $M_{\text{Bc}}$ given in Eq. (\ref{eq:MBc}). The values near $\eta = \pm 0.2$ in (b) should be understood in the context of the discussion below Eq. (\ref{eq:Mee0rephasedreal}). That is, the allowed values $(\ph,\eta) = (0,\pm(\pi+0.2))$ are phenomenologically equivalent to the values $(\ph,\eta) = (\pi,\pm0.2)$. In contrast, the values $(\ph,\eta) = (0,\pm 0.2)$ are not compatible with the data in Eqs. (\ref{eq:Vexp}) and (\ref{eq:Rexp}).}} 
	\end{minipage}
	}
\end{center}
\end{figure}

\begin{figure}[h]
\begin{center}
\fbox{
	\begin{minipage}{15 cm}
	\vspace{-5pt}
	\subfigure[$~\!R\text{ vs. }\ph\text{ for all allowed }\eta\text{ with }M_{\mu\mu}\!=\!3$]
		{
		\includegraphics[scale=0.55]{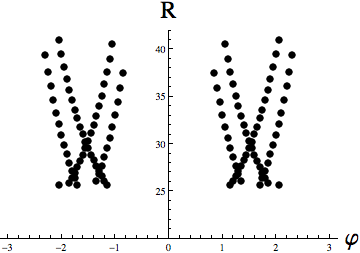}
		\label{ClassMmtEqMmm-RvsPhiForMmmEq3}
		}
	\subfigure[
	$~R\text{ vs. }\ph\text{ for all allowed }\eta\text{ with }M_{\mu\mu}\! = \!6$]
		{
		\includegraphics[scale=0.55]{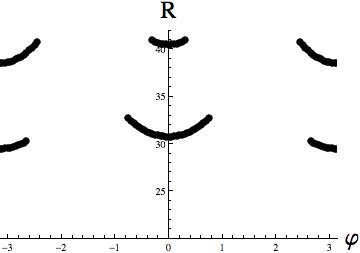}
		\label{ClassMmtEqMmm-RvsPhiForMmmEq6}
		}
		\vspace{-15pt}
	\caption{\small{The values of $R$ for the particular cases $M_{\mu\mu} = M_{\tau\tau} = |M_{\mu\tau}| = 3$ and $M_{\mu\mu} = M_{\tau\tau} = |M_{\mu\tau}| = 6$ in the matrix $M_{\text{Bc}}$ of Eq. (\ref{eq:MBc}). The angle $\eta$ ranges over all allowed values, which as shown in Fig.~\ref{ClassMmtEqMmm-PhipVsMmm} amounts to only the possibilities $\eta \sim 0.2$ (with $\ph \sim \pi$) and $\eta \sim \pi$ (with $\ph \sim 0$).}}
	\end{minipage}
	}
\end{center}
\end{figure}

\begin{figure}[h]
\begin{center}
\fbox{
	\begin{minipage}{17 cm}
	\vspace{-5pt}
	\subfigure[
	$~\!|V_{e1}|~\!\!~-~\!\!~V_{e2}\text{ vs. }\ph\text{ for all allowed }\eta\text{ with }M_{\mu\mu}~\!\!\!~=~\!\!~3$]
	{
	\includegraphics[scale=0.60]{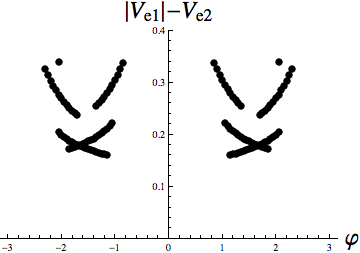}
	\label{ClassMmtEqMmm-Ve1mVe2vsPhiForMmmEq3}
	}\hspace{20pt}
	\subfigure[
	$~|V_{e1}|~\!\!~-~\!\!~V_{e2}\text{ vs. }\ph\text{ for all allowed }\eta\text{ with }M_{\mu\mu}~\!\!\!~=~\!\!~6$]
	{
	\includegraphics[scale=0.60]{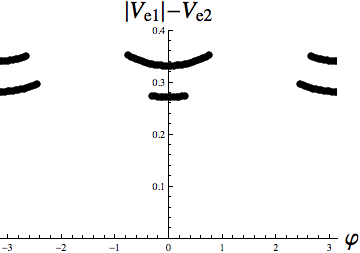}
	\label{ClassMmtEqMmm-Ve1mVe2vsPhiForMmmEq6}
	}
	\vspace{-10pt}
	\caption{\small{The values of $|V_{e1}|-V_{e2}$ for the particular cases $M_{\mu\mu} = M_{\tau\tau} = |M_{\mu\tau}| = 3$ and $M_{\mu\mu} = M_{\tau\tau} = |M_{\mu\tau}| = 6$ in the matrix $M_{\text{Bc}}$ of Eq. (\ref{eq:MBc}). Recall from Eqs. (\ref{eq:MB}) and (\ref{eq:bimax}) that in the absence of complex phases in $M_{\text{Bc}}$, the mixing matrix is of bimaximal form with $|V_{e1}|-V_{e2} = 0$, which is experimentally ruled out. The angle $\eta$ ranges over all allowed values, which as shown in Fig.~\ref{ClassMmtEqMmm-PhipVsMmm} amounts to the two possibilities $\eta \sim 0.2$ and $\eta \sim \pi$.}}
	\end{minipage}
	}
\end{center}
\end{figure}

\begin{figure}[h]
\begin{center}
\fbox{
	\begin{minipage}{16 cm}
	\vspace{-5pt}
	\subfigure[
	$~J\text{ vs. }\ph\text{ for all allowed }\eta\text{ with }M_{\mu\mu} = 3$]
	{
	\includegraphics[scale=0.60]{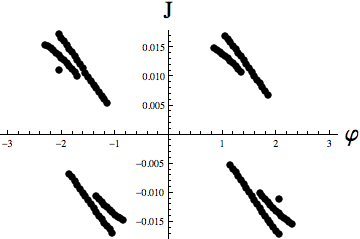}
	\label{ClassMmtEqMmm-JvsPhiForMmmEq3}
	}
	\subfigure[
	$~J\text{ vs. }\ph\text{ for all allowed }\eta\text{ with }M_{\mu\mu} = 6$]
	{
	\includegraphics[scale=0.60]{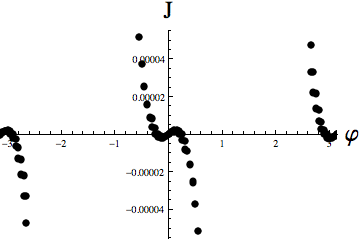}
	\label{ClassMmtEqMmm-JvsPhiForMmmEq6}
	}
	\vspace{-15pt}
	\caption{\small{The value of $J \equiv -\text{Im}(V_{e3}V_{\mu2}V_{e2}^*V_{\mu3}^*)$ for the particular cases $M_{\mu\mu} = M_{\tau\tau} = |M_{\mu\tau}| = 3$ and $M_{\mu\mu} = M_{\tau\tau} = |M_{\mu\tau}| = 6$ in the matrix $M_{\text{Bc}}$ of Eq. (\ref{eq:MBc}).}}
	\end{minipage}
	}
\end{center}
\end{figure}

We now turn to a numerical study of the matrix $M_{\text{Bc}}$ given in Eq. (\ref{eq:MBc}). Up to the phases, this is the $\mu\tau$-symmetric subcase of both Classes I and II with the additional condition $|M_{\mu\tau}| = M_{\mu\mu}$. (Note that for non-tribimaximal mixing, the phase $\ph$ is no longer necessarily zero or $\pi$.) An immediate striking feature of this matrix is given in Figure ~\ref{ClassMmtEqMmm-PhipVsMmm}, which shows that $|\eta|$ can take essentially only two possible values: $\sim 0.2$ and $\sim \pi$. This corroborates the intuition we gained from tribimaximal mixing with $|M_{\mu\tau}| = M_{\mu\mu}$.
\\
Figures ~\ref{ClassMmtEqMmm-RvsPhiForMmmEq3} and ~\ref{ClassMmtEqMmm-RvsPhiForMmmEq6} show that this CP-violating matrix interpolates between the real-valued cases with $|M_{\mu\tau}| = 4$ and $|M_{\mu\tau}| = 6$, which give $R \approx 27$ and $R = 40$ respectively. \\
\vspace{-10pt}\\
Recall that the matrix $M_{\text{Bc}}$ of Eq. (\ref{eq:MBc}) for the special case $\ph = 0$ and $\eta = \pi$ reduces to the matrix $M_{\text{B}}$ of Eq. (\ref{eq:MB}), which implies a bimaximal mixing matrix and thus $|V_{e1}| = V_{e2}$, which is incompatible with the bounds given in (\ref{eq:Vexp}). Figures ~\ref{ClassMmtEqMmm-Ve1mVe2vsPhiForMmmEq3} and ~\ref{ClassMmtEqMmm-Ve1mVe2vsPhiForMmmEq6} show that complex phases can generate mixing matrices that fall in the empirically allowed range $0.14 \leq |V_{e1}|-V_{e2} \leq 0.36$. 
\\
\begin{figure}[h]
\begin{center}
\fbox{
	\begin{minipage}{16 cm}
	\vspace{-5pt}
	\subfigure[
	$~V_{e3}\text{ vs. }\ph\text{ for all allowed }\eta\text{ with }M_{\mu\mu} = 3$]
	{
	\includegraphics[scale=0.58]{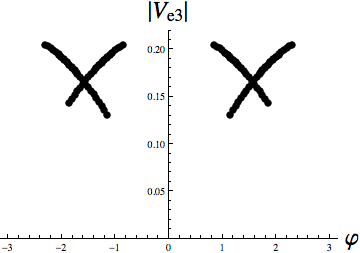}
	\label{ClassMmtEqMmm-Ve3vsPhiForMmmEq3}
	}\hspace{10pt}
	\subfigure[
	$~V_{e3}\text{ vs. }\ph\text{ for all allowed }\eta\text{ with }M_{\mu\mu} = 6$]
	{
	\includegraphics[scale=0.58]{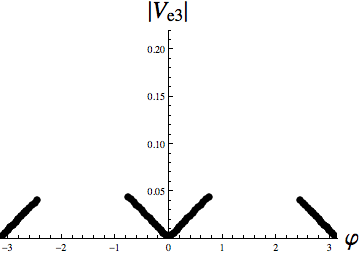}
	\label{ClassMmtEqMmm-Ve3vsPhiForMmmEq6}
	}
	\vspace{-15pt}
	\caption{\small{The value of $V_{e3}$ for the particular cases $M_{\mu\mu} = M_{\tau\tau} = |M_{\mu\tau}| = 3$ and $M_{\mu\mu} = M_{\tau\tau} = |M_{\mu\tau}| = 6$ in the matrix $M_{\text{Bc}}$ of Eq. (\ref{eq:MBc}).}}
	\end{minipage}
	}
\end{center}
\end{figure}
\\
The angular parameterization of the mixing matrix makes clear that $\del_{\text{CP}}$ only contributes to neutrino oscillations when $V_{e3}\neq 0$. Figures ~\ref{ClassMmtEqMmm-Ve3vsPhiForMmmEq3} and ~\ref{ClassMmtEqMmm-Ve3vsPhiForMmmEq6} show that the magnitude of $V_{e3}$ depends strongly on the value of $a \equiv |M_{\mu\tau}| = M_{\mu\mu} = M_{\tau\tau}$ in mass matrices with $M_{ee} = 0$ and $M_{e\mu} = |M_{e\tau}| \equiv 1$.
\\\\
More generally, amplitudes for CP-violating oscillation processes are proportional to the rephasing-invariant quantity $J \equiv -\im(V_{e3}V_{\mu2}V^*_{e2}V^*_{\mu3})$ \cite{jarlskog}. Figures ~\ref{ClassMmtEqMmm-JvsPhiForMmmEq3} and ~\ref{ClassMmtEqMmm-JvsPhiForMmmEq6} show that, like $V_{e3}$, the quantity $J$ also depends strongly on the value of $a$. For $a$ near its lower bound of $\sim 3$, the quantity $J$ is of order $\sim 10^{-2}$, but for larger $a \sim 6$ we find that $J$ is at most $\sim 10^{-5}$ and can drop to zero. For an example with a large $|V_{e3}|$ and nonzero $\del_{\text{CP}}$, and hence a large $J$, we find:
\[M = \ml 0&1&e^{\,i1.1}\\ &3&3\,e^{\,i0.2}\\ &&3 \mr \implies R \approx 39 \;\;\text{ and }\;\; |\VPMNS| = \left(
\begin{array}{ccc}
 0.79 & 0.58 & 0.19 \\
 0.52 & 0.50 & 0.70 \\
 0.32 & 0.65 & 0.69
\end{array}
\right)\]
with $\del_{\text{CP}} \approx -0.38$ and $J \approx 1.6\times10^{-2}$. In contrast, by increasing $|M_{\mu\tau}| = M_{\mu\mu} = M_{\tau\tau}$ to 6 we find
\[M = \ml 0&1&e^{\,i2.8}\\&6&6\,e^{\,i0.2}\\&&6 \mr \implies R \approx 39\;\;\text{ and }\;\; |\VPMNS| = \left(
\begin{array}{ccc}
 0.83 & 0.55 & 0.02 \\
 0.40 & 0.58 & 0.71 \\
 0.38 & 0.60 & 0.71
\end{array}
\right)\]
with $\del_{\text{CP}} \approx -2.7\times10^{-3}$ and $J \approx 1.3\times10^{-5}$. These two examples were chosen intentionally to yield the same value for $R$. Note that for $|M_{\mu\tau}| = M_{\mu\mu} = M_{\tau\tau} = 6$, keeping $\ph = 1.1$ would result in $R \approx 64$ with $|V_{e1}| \approx 0.90$, both of which are too large, and in $V_{e2} \approx 0.43$, which is too small. On the other hand, the other entries in $\VPMNS$ would all stay within the empirically allowed ranges.
\\\\
In summary, we learn that for the matrix $M_{\text{Bc}}$ decreasing the value of $M_{\mu\mu}/M_{e\mu}$ increases the value of $|V_{e3}|$ and thereby results in the possibility for larger amplitudes for CP-violating processes.
\pagebreak
\section{Discussion}\label{end}
We have studied three types of neutrino mass matrices in the flavor basis with $M_{ee} = 0$. The first two types are the CP-conserving matrices of Class I $(M_{e\mu} = M_{e\tau})$ and of Class II $(M_{\mu\mu} = M_{\mu\tau})$, which we display again for the convenience of the reader (see (\ref{eq:ClassI}) and (\ref{eq:ClassII})):
\begin{equation}\label{eq:Classes}
M_{\text{I}} \equiv \ml 0&1&1\\&M_{\mu\mu}&M_{\mu\tau}\\&&M_{\tau\tau} \mr\;\; \text{ and } \;\; M_{\text{II}} \equiv \ml 0&M_{e\mu}&M_{e\tau}\\&5&M_{\mu\tau}\\&&5 \mr
\end{equation}
The intersection of these two classes is the $\mu\tau$-symmetric ansatz (see (\ref{eq:mutau}))
\begin{equation}
M_{\mu\tau\text{-sym}} \equiv \ml 0&1&1\\&a&b\\&&a \mr
\end{equation}
which as discussed should be thought of as a useful phenomenological starting point.
\\\\
The salient phenomenological distinction between Classes I and II is that mass matrices of Class I can accommodate only a small $V_{e3}$ up to $\sim 0.03$ (Figs.~\ref{ClassI-Ve3Mmm} and~\ref{ClassI-Ve3Mtt}), while mass matrices of Class II can predict an arbitrarily large $V_{e3}$ (Figs.~\ref{ClassII-Ve3Mem} and~\ref{ClassII-Ve3Met}). Thus fundamental theories which predict a neutrino mass matrix with $M_{\mu\mu} \sim M_{\tau\tau}$ and either $M_{e\mu} \ll M_{e\tau}$ or $M_{e\mu} \gg M_{e\tau}$ (as opposed to $M_{e\mu} \sim M_{e\tau}$) will be the most constrained by future measurements of $V_{e3}$.
\\\\
The third type of matrix we studied is the complex matrix (see (\ref{eq:MBc}))
\begin{equation}
M_{\text{Bc}} \equiv \ml 0&1&e^{\,i\ph}\\&a&a\,e^{\,i\eta}\\&&a \mr\;.
\end{equation}
For this matrix, smaller values of $a$ result in larger values of $V_{e3}$ and $J \equiv -\text{Im}(V_{e3}V_{\mu2}V_{e2}^*V_{\mu3}^*)$, and thus provide experimentally promising signals of CP violation in neutrino oscillations (Figs.~\ref{ClassMmtEqMmm-Ve3vsPhiForMmmEq3} and~\ref{ClassMmtEqMmm-JvsPhiForMmmEq3}). In contrast, larger values of $a$ drive $V_{e3}$ and $J$ to zero (Figs.~\ref{ClassMmtEqMmm-Ve3vsPhiForMmmEq6} and~\ref{ClassMmtEqMmm-JvsPhiForMmmEq6}).
\\\\
A particularly interesting example is obtained from $M_{\text{Bc}}$ for the particular case $\ph = 0$ with
\begin{equation}
a = \frac{1}{2}\sqrt{m_3^2+1}\;\;\;\; \text{ and }\;\;\;\; \eta = \pi+\tan^{-1}\left( \frac{2m_3}{m_3^2-1} \right)\;.
\end{equation}
Here $m_3$ is the mass of the heaviest neutrino in units of the lighest neutrino ($m_3 > m_2 > m_1$), and the other two masses are $(m_1,m_2) = (1,2)$. In this case the mixing matrix is exactly tribimaximal, even though $M_{\mu\mu} = M_{\tau\tau} = |M_{\mu\tau}|$. (See Section \ref{complexTB}.)
\pagebreak\\
\textit{Acknowledgments:}
\\\\
This work was completed while the authors were visiting the Academia Sinica in Taipei, Republic of China, whose warm hospitality is greatly appreciated. We thank Rafael Porto for early discussions. Y.B. would like to thank Benson Way for helpful discussions. This research was supported by the NSF under Grant No. PHY07-57035.

\end{document}